\documentclass[12pt]{article}

\usepackage{amsmath,amsfonts,amssymb,latexsym}

\newtheorem{theorem}{Theorem}[section]

\numberwithin{equation}{section}
\def\be{\begin{equation}}
\def\ee{\end{equation}}
\def\bq{\begin{eqnarray}}
\def\eq{\end{eqnarray}}
\def\beq{\begin{eqnarray*}}
\def\eeq{\end{eqnarray*}}

\def\a{\alpha}
\def\b{\beta}

\begin{document}
\begin{titlepage}
\begin{flushright}
%add
\end{flushright}

\vspace{1cm}

\begin{center}
{\huge Future Singularities of Isotropic Cosmologies}

\vspace{1cm}

{\large Spiros Cotsakis$\dagger$ and Ifigeneia Klaoudatou$\ddagger$}\\

\vspace{0.5cm}

{\normalsize {\em Research Group of Cosmology, Geometry and
Relativity}}\\ {\normalsize {\em Department of Information and
Communication Systems Engineering}}\\ {\normalsize {\em University
of the Aegean}}\\ {\normalsize {\em Karlovassi 83 200, Samos,
Greece}}\\ {\normalsize {\em E-mails:} $\dagger$
\texttt{skot@aegean.gr}, \texttt{$\ddagger$ iklaoud@aegean.gr}}
\end{center}

\vspace{0.7cm}

\begin{abstract}
\noindent We show that globally and regularly hyperbolic future
geodesically incomplete isotropic universes, except for the
standard all-encompassing `big crunch', can accommodate
singularities of only one kind, namely, those having a
non-integrable Hubble parameter, $H$. We analyze several examples
from recent literature which illustrate this result and show that
such behaviour may arise in a number of different ways.  We also
discuss the existence of new types of lapse singularities in
inhomogeneous models, impossible to meet in the isotropic ones.
\end{abstract}

\vspace{1cm}
\begin{center}
{\line(5,0){280}}
\end{center}

\end{titlepage}

\section{Introduction}
It has been pointed out very recently that finite-time, finite
Hubble parameter singularities can occur in the expanding
direction of  even the simplest FRW universes  in general
relativity and other metric theories of gravity and, although such
behaviour depends on the details of the particular model, yet, it
is met in a very wide and extremely varied collection of possible
cosmologies (cf. Refs \cite{cald99}-\cite{add2} for a partial
list). These singularities cannot be accommodated by the usual
singularity theorems \cite{ha-el73}, for their features arise from
\emph{necessary}, not sufficient, conditions for their existence
while their character is somewhat milder than the standard,
all-encompassing `big crunch' singularities met in other model
universes. This situation appears somewhat wanting as all known
relevant models provide merely examples, not general theorems, of
such a phenomenon.

What is the underlying reason for such a behaviour?
In this paper, we show that the finite-time, finite-$H$
singularities met in isotropic cosmologies
satisfy all the assumptions of the completeness theorem of \cite{ch-co02}
\emph{except} the one about an infinite proper-time interval of existence of
privileged observers and because of this reason are geodesically incomplete.
This leads to a classification of the possible types of singularities in the
isotropic category based on necessary analytic conditions for
their existence. We also briefly discuss how new types of such
singularities may  arise in more general inhomogeneous models.

\section{Necessary conditions for future singularity formation}
Although we shall focus  exclusively on isotropic models,  it is
 instructive to begin our analysis by taking a more
general stance. Consider a spacetime $(\mathcal{V},g)$ with
$\mathcal{V}=\mathcal{M}\times \mathcal{I},\;$ $\mathcal{I}
=(t_{0},\infty )$, where
$\mathcal{M}$ is a smooth manifold of dimension $n$ and
$^{(n+1)}g$ a Lorentzian metric which in the usual $n+1$
splitting, reads
\begin{equation}
^{(n+1)}g\equiv -N^{2}(\theta ^{0})^{2}+g_{ij}\;\theta ^{i}\theta ^{j},\quad
\theta ^{0}=dt,\quad \theta ^{i}\equiv dx^{i}+\beta ^{i}dt.  \label{2.1}
\end{equation}
Here $N=N(t,x^{i})$ is called the \emph{lapse function}, $\beta
^{i}(t,x^{j})$ is called the \emph{shift function} and the spatial
slices $\mathcal{M}_{t}\,(=\mathcal{M}\times \{t\})$ are spacelike
submanifolds endowed with the time-dependent spatial metric
$g_{t}\equiv g_{ij}dx^{i}dx^{j}$. We call such a spacetime a
\emph{sliced space} \cite{co03}. A sliced space is time-oriented
by increasing $t$ and we choose $\mathcal{I} =(t_{0},\infty )$
because we shall study the future singularity behaviour
of an expanding universe with a singularity in the past,
for instance at $t=0<t_{0}.$ However, since $ t$ is just a
coordinate, our study could apply as well to any interval
$\mathcal{I}\subset\mathbb{R}$.

A natural causal assumption for $(\mathcal{V},g)$ is that it is
\emph{globally hyperbolic}. This implies the existence of a time function
on $(\mathcal{V},g)$. In a globally hyperbolic space
the set of all timelike paths joining two points is compact in the
set of paths, and spacetime splits as above with each spacelike
slice $\mathcal{M}_{t}$ a Cauchy surface, i.e., such that each
timelike and null path without end points cuts $\mathcal{M}_{t}$
exactly once \cite{ge70}.

We say that a sliced space has \emph{uniformly bounded
lapse} if the lapse function $N$ is bounded below and above by
positive numbers $N_{m}$ and $N_{M}$,
\begin{equation}
0<N_{m}\leq N\leq N_{M}. \label{h1}
\end{equation}
A  sliced space has \emph{uniformly bounded shift} if the $g_{t}$
norm of the shift vector $\beta$, projection on the tangent space
to $\mathcal{M}_{t}$ of the tangent to the lines $\{x\}\times
\mathcal{I}$, is uniformly bounded by a number $B.$

A  sliced space has \emph{uniformly bounded spatial metric} if the
time-dependent metric $g_{t}\equiv g_{ij}dx^{i}dx^{j}$ is
uniformly bounded below for all $t\in \mathcal{I}$ by a
metric $\gamma =g_{t_0}$, that is there exists a number $A>0$ such
that for all tangent vectors $v$ to $\mathcal{M}$ it holds that
\begin{equation}
A\gamma _{ij}v^{i}v^{j}\leq g_{ij}v^{i}v^{j}.  \label{h2}
\end{equation}
A sliced space $(\mathcal{V},g)$ with uniformly bounded
lapse, shift and spatial metric is called \emph{regularly hyperbolic}.

Denoting by $\nabla N$ the space gradient of the lapse $N$, by
$K_{ij}=-N \Gamma_{ij}^{0}$ the extrinsic curvature of
$\mathcal{M}_{t}$, and by $|K|^2_{g}$ the product
$g^{ai}g^{bj}K_{ab}K_{ij}$, we have the following theorem which
gives sufficient conditions for geodesic completeness
\cite{ch-co02}:
\begin{theorem}\label{1}
Let  $(\mathcal{V},g)$ be a sliced space such that the following
assumptions hold:
\begin{description}
\item[C1] $(\mathcal{V},g)$ is globally hyperbolic
\item[C2] $(\mathcal{V},g)$ is regularly hyperbolic
%(i.e., $N$, $\beta$ and $g_{t}$   uniformly bounded)
\item[C3] For each finite $t_{1},$ the space gradient of the lapse, $|\nabla N|_{g}$,
is bounded by a function of $t$ which is integrable on $[t_{1},+\infty )$
\item[C4] For each finite $t_{1},$  $|K|_{g}$ is bounded by a function of $t$ which is
integrable on $ [t_{1},+\infty )$.
\end{description}
Then  $(\mathcal{V},g)$ is future timelike and null geodesically
complete.
\end{theorem}
It is known that in a regularly hyperbolic spacetime,
condition $C1$ is in fact \emph{equivalent} to
the condition that each slice of $(\mathcal{V},g)$ is a complete
Riemannian manifold, cf. \cite{co03}. The completeness theorem (\ref{1}) gives
sufficient conditions for timelike and null geodesic completeness
and therefore implies that the negations of each one of conditions $C1-C4$
are \emph{necessary} conditions for the existence of
singularities (while, the singularity
theorems (cf. \cite{ha-el73}) provide sufficient conditions for this
purpose). Thus Theorem \ref{1} is precisely what is needed for the analysis of
models with a big rip singularity
inasmuch these models require \emph{necessary} conditions for future
singularities.

Now for an FRW metric,   $N=1$ and $\beta =0$, and so condition $C2$
is satisfied provided the scale factor $a(t)$ is a bounded from below function of the
proper time $t$ on $\mathcal{I}$. Condition $C3$ is trivially
satisfied and this is again true in more general
homogeneous cosmologies where the lapse function $N$ depends only
on the time and not on the space variables. In all these cases its
space gradient, $|\nabla N|_{g}$, is zero. Condition $C4$ is the
only other condition which can create a problem. For an isotropic universe
 ${|K|_{g}} ^{2}=3({\dot{a}/a})^{2}=3H^{2}$, and
so we conclude that FRW universes in which  the scale factor is
bounded below can fail to be complete only when $C4$ is violated.
This can happen in only one way: There is a finite time $t_1$ for
which $H$ fails to be integrable on the time interval
$[t_1,\infty)$. Since this non-integrability of $H$ can be
implemented in different ways, we arrive at the following result
for the types of future singularities that can occur in isotropic
universes.
\begin{theorem}\label{2}
Necessary conditions for the existence of future singularities in
globally hyperbolic, regularly hyperbolic FRW universes are:
\begin{description}
\item[S1] For each finite $t$, H is non-integrable on $[t_1,t]$,
or
\item[S2] H blows up in a finite time, or
\item[S3] H is defined and integrable (that is bounded, finite) for only a finite
proper time interval.
\end{description}
\end{theorem}
When does Condition $S1$ hold? It is well known that a function
$H(\tau)$ is integrable on an interval $[t_1,t]$ if $H(\tau)$ is
defined on $[t_1,t]$, is continuous on $(t_1,t)$ and the limits
$\lim_{\tau\rightarrow t_1^+}H(\tau)$ and $\lim_{\tau\rightarrow
t^-}H(\tau)$ exist. Therefore there are a number of different ways
which can lead to a singularity of the type $S1$ and such
singularities are  in a sense more subtle than the usual ones
predicted by the singularity theorems. For instance,   they may
correspond to `sudden' singularities (see \cite{ba04} for this
terminology) located at the right end (say $t_s$) at which  $H$ is
defined and finite but \emph{the left limit},
$\lim_{\tau\rightarrow t_1^+}H(\tau)$, may fail to exist, thus
making $H$ non-integrable on $[t_1,t_s]$, for \emph{any} finite
$t_s$ (which is of course arbitrary but fixed from the start). We
shall see examples of this behaviour in the next Section.
Condition $S2$ leads to what is called here a blow-up singularity
corresponding  to a future singularity characterized by a blow-up
in the Hubble parameter. Note that $S1$ is not implied by $S2$ for
if $H$ blows up at some finite time $t_s$ after $t_1$, then it may
still be integrable on $[t_1,t]$, $t_1<t<t_s$. Condition $S3$ also
may lead to a big-rip type singularity, but for this to be a
genuine type of singularity (in the sense of geodesic
incompleteness) one needs to demonstrate that  the metric is
non-extendible to a larger interval.

\section{Examples}
We give below some representative examples to illustrate the
results of the previous Section. An example of a blow-up
singularity  is the recollapsing flat FRW model filled with dust
and a scalar field with a `multiple' exponential potential
considered in \cite{ga04}. These authors choose the potential to
be of the form $V(\phi)=W_{0}-V_{0}\sinh(\sqrt{3/2}\kappa \phi)$
where $W_{0}$ and $V_{0}$ are arbitrary constants and
$\kappa=\sqrt{8\pi G_{N}}$, and split the scale factor according
to the transformation ${a}^{3}=xy$. Here ones sets
$x=C[\exp\chi_1\cos\chi_{2}+\exp (-\chi_{1})\cos\chi_{2}]$,
$y=C[\exp\chi_1\sin\chi_{2}+\exp (-\chi_{1})\sin(-\chi_{2})]$ with
$\chi_1 =w_{1}(t-t_{0}), \chi_{2}=w_{2}(t-T_{0})$, where
 $C>0$, $t_{0}$ an arbitrary constant, $T_{0}$ is the
`initial' time and $w_{1}$, $w_{2}$ positive parameters such that
$w_{1}^{2}-w_{2}^{2}=3/4{\kappa}^{2}W_{0},\,\,
2w_{1}w_{2}=3/4{\kappa}^{2}V_{0}.$ Using the results of Section 2, we can immediately see why
this model has such a singularity (in \cite{ga04} this was proved
by different methods). For large and positive values of the time parameter $t$ the scale
factor becomes $
a=C^{2/3} \exp (2/3 \chi_{1})\cos ^{1/3}
\chi_{2}\sin^{1/3}\chi_{2},
$
which is obviously divergent.
We then find ${|K|_{g}} ^{2}$ to be essentially (that is
except unimportant constants) proportional to $w_{2}(\cot\chi_2 -\tan\chi_2
)/3$,
and so the extrinsic curvature blows up as  $t$ tends to the \emph{finite} time value
$\pi/(2w_{2})+T_{0}$. This is a blow-up singularity in the sense
of Condition $S2$ of Section 2.

Another example of a future blow-up singularity is provided by
the `phantom' cosmologies (see relevant references in
\cite{ba04}). In all these models different methods are used,
depending each time on the analysis of the field equations of each
phantom model, to prove the existence of such future
singularities. For instance, in \cite{go_2 04}
we meet a flat FRW universe with a minimally coupled
scalar field $\phi$ in Einstein's gravity  with the equation of state
$p=w\rho$, with $w<-1$ (phantom dark energy). In this model the scale
factor takes the form
\begin{equation}
a=\left[{a _{0}}^{3(1+w)/2}+\frac {3(1+w)\sqrt
{A}}{2}(t-t_{0})\right]^{\frac {2}{3(1+w)}},
\end{equation}
where $A=8\pi G C/3$ and $C$ an integration constant. Hence the extrinsic curvature becomes
\begin{equation}
|K|_{g}^{2}=3A\left[{a _{0}}^{3(1+w)/2}+\frac {3(1+w)\sqrt
{A}}{2}(t-t_{0})\right]^{-2},
\end{equation}
and so when $w<-1$ and $t=t_{0}+2/[3\sqrt {A}(|w|-1)]a_{0}^{3(1-|w|)/2}$,
the extrinsic curvature diverges thus  causing a blow-up
singularity.
Similar behaviour is found in other singular phantom cosmologies,
for instance those of Refs. \cite{cald99}, \cite{cald03},
\cite{yu03}, \cite{go03}, \cite{ch04}, namely,  there is \emph{some} finite
time $t_{f}$ at which the Hubble parameter blows up.

Had the evolution  was characterized by an integrable $H$
\emph{for every $t$}, then we would expect from Theorem \ref{1}  all these
models to be timelike and null geodesically complete. An example
illustrating this is given by the  flat FRW universe filled with phantom dark energy
which behaves simultaneously as Chaplygin gas studied in \cite{sr04} (see also \cite{go_1 04}).
Here the phantom dark energy component satisfies the equation of state
$p=w\rho$ with $ w<-1,$ as well as the equation of state of a Chaplygin gas
$p=-A/\rho,$ where $A$ is a positive constant (similar results will hold when
$w\in (-1,-1/3)$ -$k$-essence
models, see, e.g., \cite{go_2 04}). Then integrating  the continuity equation
$
\dot{\rho}=-3\dot{a}(\rho+p)/a,
$
we find
\begin{equation}\label{8.4}
\rho^{2}(t)=A+(\rho_{0}^{2}-A){(a_{0}/a(t))}^{6}.
\end{equation}
or, using the two equations of state and substituting to (\ref{8.4}) leads to
\begin{equation}
\rho(t)={\rho_{0}}^{2}{[-w_{0}+(1+w_{0}){(a_{0}/a(t))}^{6}]}^{1/2}
\end{equation}
where $A=-w_{0}{\rho_{0}}^{2}$, $w_{0}<-1$. Finally, the solution
for the scale factor reads
$
a(t)=Ce^{C_{1}(t-t_{0})},
$
and so the Hubble parameter in this case is constant,
$
H=C_{1}/6.
$
This non singular behavior is due
to the fact that all the conditions of the Completeness Theorem
\ref{1} are met.

We now move on to an example of a  big rip singularity in the
sense Condition $S1$ of Section 2. Barrow in \cite{ba04} rightly
calls such singularities `sudden'. He considers an expanding FRW
model with a fluid that satisfies the energy conditions $\rho >0$
and $\rho +3p>0$, and shows that the special solution
\begin{equation}
a(t)=1+\left(\frac{t}{t_{s}}\right)^{q}
(a_{s}-1)+(1-\frac{t}{t_{s}})^n,
\end{equation}
with $1<n<2, 0<q\leq 1$ and $a_s\equiv a(t_s)$, exists, as a
smooth solution, only on the interval $(0,t_{s})$, while $a_{s}$
and $H_{s}\equiv H(t_s)$ are finite at the right end. Note that
setting $A=q(a_s-1)/t^q_s, B=n/t_s^n$, we find that
$\dot{a}=At^{q-1}+B(t_s-t)^{n-1}$ which means that, unless $q=1$,
$\dot{a}$ blows up as $t\rightarrow 0$, making $H$ continuous only
at $(0,t_s)$. Also $a(0)$ is finite and we can extend $H$ and
define it to be finite also at $0$, $H(0)\equiv H_0$, so that $H$
is defined on $[0,t_s]$). However, since $\lim_{t\rightarrow
0^+}H(t)=\pm\infty$,  we conclude that this model universe
implements exactly Condition $S1$ of the previous Section and thus
$H$ is non-integrable on $[0,t_s]$, $t_s$ arbitrary. This is a big
rip singularity characterized by the fact that as $t\rightarrow
t_{s}$ one obtains $\ddot{a}\rightarrow-\infty$. Then using the
field equation we see that this is really a divergence in the
pressure, $p/2\rightarrow-\ddot {a}/{a}-H^{2}/{2}-k/(2a^{2})$, and
so $p\rightarrow\infty$. In particular, we cannot have in this
universe a family of privileged observers each having an infinite
proper time and finite $H$.

Another way to see this result in the particular case when $\rho > 0$ and $p\geq 0$,
is to use the following result from \cite{ha-el73}, pp. 141-2:
In an FRW universe with $\rho > 0$ and $p\geq 0$, given any vector $X$ at any point
$q$,
the geodesic $\gamma (\upsilon)$ which passes through the point $q=\gamma(0)$ in the
direction of $X$, is such that either
\begin{itemize}
 \item $\gamma (\upsilon)$ can be extended to arbitrary
values of its affine parameter $\upsilon$, \emph{or}
\item there is some value $\upsilon_{0}>0$
such that the scalar invariant $(R_{\a\b}-1/2
Rg_{\a\b})(R^{\a\b}-1/2Rg^{\a\b})$ is unbounded on $\gamma
([0,\upsilon])$.
\end{itemize}
Therefore under the assumptions of \cite{ba04}, the invariant $(R_{\a\b}-1/2
Rg_{\a\b})(R^{\a\b}-1/2Rg^{\a\b})$ is calculated to be
\begin{equation}
\frac{12}{{a}^{4}}+\frac{24\dot{a}^{2}}{a^{4}}+\frac{12\dot{a}^{4}}{a^{4}}+\frac{12\ddot
a}{a ^{3}}+\frac{12\dot{a} ^{2}\ddot{a}}{a^{3}}+\frac{12\ddot{a}^{2}}{a^{2}},
\end{equation}
and since $a\rightarrow a (t_{s})$, $ H(t)\rightarrow H_{s}$,
$p(t)\rightarrow \infty $, $\ddot{a}/ a \rightarrow -\infty$ as $t
\rightarrow t_{s}$, we see that $(R_{\a\b}-1/2
Rg_{\a\b})(R^{\a\b}+1/2Rg^{\a\b})$ is unbounded at $t_s$. Hence we
find that this spacetime is geodesically incomplete.

Further, Carroll \emph{et al} in \cite{ca03} provide a different
model leading, however, to a similar future big rip singularity.
They start from a theory of gravity with lagrangian density $L=R+\alpha
R^{-2} +L_{mat}$ and conformally transform to an Einstein frame to
get an equivalent theory which is described as general relativity
coupled to a system comprised by the scalar field resulting from
the conformal transformation and the conformally transformed
matter. Since the two matter components in the Einstein frame are
not separately conserved, there is now a non-trivial scalar
field-matter interaction which manifests itself with new terms
appearing  in the field equations. It then follows that as
$\phi\rightarrow 0$, the time derivative of the extrinsic
curvature blows up, $\dot{H}\rightarrow\infty$, while $H$ itself
is finite at $\phi =0$ (cf. \cite{ca03}). This is  clearly a big-rip type
singularity.

A different example of a big rip singularity is given by
Borde \emph{et al} in \cite{bo01}. They  showed that in
inflating  spacetimes just a bounded averaged-out Hubble function
is necessary to produce a singularity with \emph{finite} $H$. Why
do they obtain such a behaviour? To answer this question we again
use our Theorem \ref{2} and conclude that their condition given by
Eq. (11) in \cite{bo01} is precisely the negation of condition
$C4$ of Theorem \ref{1} above. In effect, what these authors do is
to find that a necessary condition for a singularity in the
spacetimes considered is that $H$ be finite on a \emph{finite}
time interval. One recognizes that under their assumptions,
privileged observers cannot exist in inflating universes for an
infinite proper time, because had such observers existed, Theorem
\ref{1} would then imply that these inflating spacetimes are
timelike and null geodesically complete. More precisely, in \cite{bo01}
one starts with a flat FRW  model,
$
ds^{2}=dt^{2}-a^{2}(t)d\bar{x}^{2},
$
and considers all quantities along a null geodesic with affine parameter $\lambda$.
Since the model is conformally Minkowski we have
$
d\lambda \propto a(t) dt,
$
or,
$
d\lambda=a(t)dt/a(t_{s}),
$
so that $d\lambda/dt=1$ for $t=t_{s}$, where $t_{s}$ is a finite
value of time. One then defines
\begin{equation}
H_{av}=\frac{1}{\lambda(t_{s})-\lambda(t_{i})}
\int_{\lambda(t_{i})}^{\lambda(t_{s})}H(\lambda)d\lambda,
\end{equation}
and so  if $H_{av}>0$ one finds
\begin{equation}\label{7.5}
0<H_{av}=\frac{1}{\lambda(t_{s})-\lambda(t_{i})}
\int_{\lambda(t_{i})}^{\lambda(t_{s})}H(\lambda)d\lambda
=\frac{1}{\lambda(t_{s})-\lambda(t_{i})}
\int_{a(t_{i})}^{a(t_{s})}\frac{da}{a(t_{s})}\leq\frac{1}
{\lambda(t_{s})-\lambda(t_{i})}.
\end{equation}
This shows that the affine parameter must take values only in a finite interval
which implies geodesic
incompleteness. A similar proof is obtained for the case of a timelike
geodesic. Notice that condition (\ref{7.5}) holds if and only if
condition $C4$ of Theorem \ref{1} is valid only for a finite interval of time, thus
leading to incompleteness according to Theorem \ref{2}.
A similar bound for the Hubble parameter is obtained in \cite{bo01}  for the general
case, and one therefore concludes that such a model must be
geodesically incomplete.

A relaxation of the requirement  that  $H$ be finite for only a
finite amount of proper time leads,  as expected by Theorem \ref{1},
to singularity-free inflationary models evading the previously
encountered singularity behaviour. Such models have been
considered in \cite{el-ma04} (see also \cite{ag01}, \cite{ag03}).
It is particularly interesting in our context to see why such
behaviour actually occurs: In \cite{el-ma04}, the scale factor is
assumed to be bounded below by a constant $a_{i}$ (so that this
universe is regularly hyperbolic in our notation), and  its
form is given by (these authors also take this to
be qualitatively true in more general models),
\be
a(t)=a_{i}\left[ 1+\exp(\sqrt{2}t/a_{i})\right]^{1/2}.
\ee
One then  finds that $H$ is not only bounded as
$t\rightarrow -\infty$, but actually becomes zero asymptotically. Thus by Theorem
\ref{1} this universe is geodesically complete. It is instructive, however, to
notice  that a
slight change in the inflationary solution can drastically alter
this behaviour. Consider for example the old quasi-exponential models
first considered by Starobinski  (see  \cite{co-fl93} and references therein), where the scale
factor is given by $a=c_{0}\exp\left[{c_{1}t+c_{2}t^2}\right]$, with the
$c_{i}$'s constants. Then we see that $H=c_{1}+2c_{2}t$ and so it
blows up to $\pm\infty$ as $t\rightarrow\pm\infty$ depending on the sign of the constant
$c_{2}$. It is interesting to further notice (cf. \cite{co-fl93}) that this solution
is an attractor of all
homogeneous, isotropic solutions of higher order gravity theories
and so we may conclude that the blow up behaviour leading to a singular regime is perhaps a
more generic feature than completeness in such contexts.

Examples of  big rip as well as geodesically complete types of
evolution are also met in many brane cosmologies, cf. \cite{ss1},
\cite{ss2}, \cite{ss3}, \cite{tam03}, \cite{fo03}, \cite{br04}.
The singularities in such models are all characterized by the fact
that there exists no admissible slicing  with an infinite proper
time of privileged observers and the Hubble parameter remains
finite but only for this finite interval of proper time  and
cannot be defined beyond that interval (cf. \cite{ss1}), and hence
in accordance with Theorem \ref{2} these spacetimes must be
geodesically incomplete.

However, geodesically complete solutions also exist and such
phases in brane evolution are described in  detail by the series
of brane models studied in \cite{tam03}. The main idea is rather
simple, cf. \cite{ss3}, \cite{br04}. We start with a flat 3-brane
filled with ordinary matter as well as phantom dark energy
embedded in a 5- dimensional bulk. This model goes through a
series of cycles of finite accelerating expansion and contraction.
At the end of each cycle it bounces avoiding in this way a
singular behavior. In the expansion phase  the phantom dark energy
component increases and drives cosmic acceleration. This rapid
acceleration tears apart any bound structure produced during
expansion and towards the end of the expanding phase phantom dark
energy has become so high that modifications of the Friedman
equation become important. The modified Friedman equation on the
$3$-brane is
\begin{equation}
H^{2}=\frac{\Lambda_{4}}{3}+\frac{8\pi}{3M_{p}^{2}}\rho+
\epsilon({\frac{4\pi}{3M_{5}^{3}}})^{2}\rho
^{2}+\frac{C}{a^{4}},
\end{equation}
where C is an integration constant, $\epsilon$ corresponds to the
metric signature of the extra dimension and $\Lambda_{4}$ is the
cosmological constant on the brane.
Assuming that $C/a^{4}$ is negligible and $\epsilon<0$, we have
for the critical brane case ($\Lambda_{4}$=0)
\begin{equation}\label{3.10}
H^{2}=\frac{8\pi}{3M_{p}^{2}}(\rho-\frac{\rho^{2}}{2|\sigma|}),
\end{equation}
hence, $H=0$ when $\rho_{bounce}=2|\sigma|$ where $\sigma$ is
the tension of the brane.
At this scale the model will turn around and start to contract.
During contraction the phantom component will decrease but at
late times, when the model becomes matter or radiation dominated,
the energy density will be very high and modifications of the
Friedman equation will again become important. When the critical
value of the density is reached the model will bounce and start
to expand. Using the assumptions of this model, we
 see from Eq. (\ref{3.10}) that the Hubble parameter remains eternally
finite and also all other assumptions of Theorem \ref{1} are satisfied.
Therefore this universe is geodesically complete.

\section{Discussion}
The main result of this paper is the recognition that future big
rip singularities occur because there is a finite time, say at
$0$, such that the Hubble parameter is not integrable on
$[0,\infty)$. In such universes privileged observers cannot exist
for an infinite proper time starting from $0$. We know from
Theorem \ref{1} (cf. \cite{ch-co02}) that if such observers
existed for an infinite proper time, then such a universe would be
timelike and null geodesically complete. This condition is not
satisfied in recent models with big rip singularities and
therefore such universes may not be complete. In fact, examples
show, cf. Refs. \cite{bo01}, \cite{ca03}, \cite{ba04}, that such
spaces are necessarily singular.

It is interesting that isotropic models coming from completely
different motivations reveal their tendencies to have future
singularities (that is future geodesic incompleteness) for exactly
one the same reason, namely the non-integrability of $H$. Of
course, the fact that this condition can be sustained in different
models stems from the different ways these models are constructed
and the specific features they share. It may be supported by the
presence of a scalar field inducing a novel interaction with
matter as in the case of $f(R)$ theory considered in Ref.
\cite{ca03}, or lead to a divergence in the fluid pressure and a
corresponding one in $\ddot{a}$ as in \cite{ba04}, or in the
inflationary character of the particular model as in \cite{bo01},
etc.

We have further shown that when we restrict attention to
homogeneous and isotropic universes,  these non-integrable-$H$
singularities are the only types of future big-rip singularities
which occur. This leads us to the following question: Is it
possible to find a globally hyperbolic, regularly hyperbolic
inhomogeneous cosmology which satisfies condition $C4$ of Theorem
\ref{1} but \emph{not} $C3$? Since $|\nabla N|_{g}$ is now a
function not only of the time but also of the space variables it
is not automatically zero. If true, this effect will lead, in
addition to the ones already present in the isotropic case (i.e.,
those having an integrable $|\nabla N|_{g}$ but failing $C4$),  to
two new types of \emph{lapse singularities} both not satisfying
condition $C3$. Namely, that which has a diverging $|\nabla
N|_{g}$ in a finite time corresponding to a \emph{blow-up lapse
singularity} and secondly, that with $|\nabla N|_{g}$ finite only
for a finite interval of proper time, a \emph{big-rip lapse
singularity}. The latter singularity, if it exists, will
necessarily have several qualitative features distinct from the
corresponding blow-up or big-rip (extrinsic) curvature
singularities discussed here. In turn, these will not anymore be
the only available types of singularity in the inhomogeneous
category, as indeed they are for isotropic universes.

\section*{Acknowledgement}
This paper considers the application to cosmological models of
joint previous work of Y. Choquet-Bruhat and the first author and
we are deeply indebted to her for many useful comments. We are grateful to
S. Nojiri, S. Odintsov, A. Starobinski and A. Vikman for useful
exchanges and communications. This work
was supported by the joint Greek Ministry of Education/European Union Research
grants `Pythagoras' No. 1351, `Heracleitus' No. 1337, and by a State Scholarship
Foundation grant, and this support is gratefully acknowledged.

\end{document}